\documentclass[a4paper,11pt]{article}

\usepackage{fullpage}
\usepackage{amsmath}
\usepackage{amssymb}
\usepackage{stmaryrd}
\usepackage{graphicx}
\usepackage{accents}

\newcommand*{\bdot}[1]{\accentset{\mbox{\large\bfseries .}}{#1}}

\title{On the general governing equations of electromagnetic acoustic transducers}

\author{Prashant Saxena \\ \small{Chair of Applied Mechanics, University of Erlangen-Nuremberg,} \\ \small{Egerlandstrasse 5, 91058 Erlangen, Germany} \\ \small{Email: prashant.saxena@ltm.uni-erlangen.de}}
\date{}
\begin{document}
\maketitle

 \begin{abstract}
 In this paper, we present the general governing equations of electrodynamics and continuum mechanics that need to be considered while mathematically modelling the behaviour of electromagnetic acoustic transducers (EMATs). We consider the existence of finite deformations for soft materials and the possibility of electric currents, temperature gradients, and internal heat generation due to dissipation. Starting with Maxwell's equations of electromagnetism and balance laws of nonlinear elasticity, we present the governing equations and boundary conditions in incremental form in order to solve wave propagation problems of boundary value type.
 \end{abstract}


\section{Introduction}

Electromagnetic acoustic transduction is a technique used to generate and detect mechanical waves in magnetoelastic conductors for the purpose of nondestructive testing. The devices, called electromagnetic acoustic transducers (EMATs) or electromagnetic acoustic receivers (EMARs), function by exploiting the coupling between electromagnetic and mechanical effects in a deformable continuum. 
The transducer mechanism (as shown in the schematic in Figure 1) consists of a current-carrying coil suspended on the specimen under test which is magnetized by a large static magnetic field. An AC current is passed through the coil which generates a time-varying magnetic field in the specimen. This, on interaction with the static magnetic field, leads to an electromagnetic body force that causes the generation of mechanical waves in the specimen. In the case of EMARs, the existing mechanical waves in the specimen, on interaction with the static magnetic field, produce time-varying magnetic field in the vicinity. This generates an AC current in the current-carrying coil which can be measured to analyze the mechanical waves in the specimen. Thus, it is very important to understand the exact coupling between mechanical and electromagnetic fields for an accurate mathematical modelling of the process.

Several orientations of the specimen and the coil for EMATs have been studied in the recent literature. Much of the work (including \cite{Hirao2003}) is based, among other literature, on the papers by Ludwig et al. \cite{Ludwig1993}, Ogi \cite{Ogi1997}, and Thompson \cite{Thompson1978} on the theory and numerical simulation of EMATs. Recently, Shapoorabadi et al. \cite{Shapoorabadi2005} have presented the governing equations of EMATs using a formulation involving the magnetic vector potential. The list of references is by no means exhaustive and there are many more papers and theses on this subject.

However, the existing theory of EMATs is, in our view, incomplete in some respects. Many theoretical models have been developed and experimental results been obtained for only a sinusoidal steady state of current and displacement (such as \cite{Thompson1978}). A large volume of literature assumes linear elastic deformations in the specimen and a linear coupling between the electromagnetic and mechanical phenomena (such as \cite{Ludwig1993}, \cite{Shapoorabadi2005}). This is, however, not the case in reality with the coupling being nonlinear, as has been presented in the classical texts by Pao \cite{Pao1978}, Eringen and Maugin \cite{Eringen1990}, and in the recent paper by Maugin \cite{Maugin2009}. Moreover, with the development of soft polymeric electro- and magnetoelastic smart materials (\cite{Jolly1996}, \cite{Bose2012}) in the recent years, one needs to take into account the possibility of finite deformations. In some materials, heat generation caused by electric currents may lead to significant changes in magneto-elastic properties, hence these effects also need to be taken into account.

In this paper, we present the equations that govern wave propagation in a finitely deformed solid in the presence of electromagnetic fields. On the existing finite deformation, magnetization, and electric polarization, time dependent small increments in the electric, magnetic, temperature, and deformation fields are allowed. We provide a general constitutive formulation based on a free energy function that depends on the initial deformation, underlying electromagnetic fields, and temperature. This is based on a generalization of the constitutive formulations of electroelasticity and magnetoelasticity provided by Dorfmann and Ogden \cite{Dorfmann2004,Dorfmann2006}. The same formulation of magnetoelasticity has been used by Destrade and Ogden \cite{Destrade2011}, and author and co-workers \cite{Saxena2011,Saxena2012} to study some wave propagation problems in magnetoelasticity.

In Section 2, we recapitulate the basic governing equations of electromagnetism in continua and the laws of thermodynamics as given by Pao \cite{Pao1978} and Eringen and Maugin \cite{Eringen1990}. These are used to derive constitutive relations from a free energy function. We then allow for small time-dependent increments in Section 4 to study wave propagation. Several `moduli' tensors are also introduced in this section to quantify the couplings in elastic, electric, magnetic, and temperature fields. It is expected that the general equations provided here will be solved using  numerical schemes such as FEM for specific boundary problems related to EMATs.

\section{Basic equations}

The undeformed stress-free \emph{reference configuration} of a continuous elastic body is denoted by $\mathcal B_r$ and its boundary by $\partial \mathcal B_r$. Let $\mathcal B_t$, the \emph{current configuration}, be the region occupied by the body at time $t$ and $\partial \mathcal B_t$ its boundary. The material points of body are identified by the position vector $\mathbf{X}$ in $\mathcal B_r$ which becomes the position vector $\mathbf{x}$ in $\mathcal B_t$.

\subsection{Kinematics}
The time-dependent deformation (or motion) of the body is described by an invertible mapping $\boldsymbol{\chi}$ that maps points from $\mathcal B_r$ to points in $\mathcal B_t$ such that $\mathbf{x} = \boldsymbol{\chi}(\mathbf{X},t)$. The function
$\boldsymbol{\chi}$ and its inverse are assumed to be sufficiently regular in space and time. The velocity $\mathbf{v}$ and
acceleration $\mathbf{a}$ of a material particle at $\mathbf{X}$ are defined by
\begin{equation}
\mathbf{v}(\mathbf{x}, t) = \mathbf{x}_{, t} = \frac{\partial}{\partial t} \boldsymbol{\chi} \left( \mathbf{X},t \right),
\quad \mathbf{a}(\mathbf{x},t) = \mathbf{v}_{,t} = \mathbf{x}_{,tt} = \frac{\partial^2}{\partial t^2} \boldsymbol{\chi} \left( \mathbf{X},t \right),
\end{equation}
where the subscript $t$ following a comma denotes the material time derivative. In this paper; Grad, Div, and Curl denote the standard differential operators in the reference configuration while grad, div, and curl denote the same in the current configuration.
 The \emph{deformation gradient tensor} is defined as $\mathbf{F} = \mbox{Grad}\, \boldsymbol{\chi}(\mathbf{X},t)$ and its determinant is denoted by $J = \det \mathbf{F}$, with $J>0$. Associated with $\mathbf{F}$ is the right Cauchy-Green tensor $\mathbf{c} = \mathbf{F}^{\mathrm T} \mathbf{F}$. For an incompressible material the constraint
$ J = \mbox{det}\, \mathbf{F} = 1,$
needs to be satisfied.

\subsection{Maxwell's equations}
The well-known Maxwell's equations governing electromagnetic fields in a deformable continua (the region $\mathcal B_t$ in Figure 1) are given by the Dipole-Current Circuit Model of Pao \cite{Pao1978} as
\begin{align}
\mbox{div}\, \mathbf{B} = 0, \quad \mbox{curl}\, \mathbf{E} + \frac{\partial \mathbf{B}}{\partial t} = 0, \quad \mbox{div} (\varepsilon_0 \mathbf{E} + \mathbf{P}) = \rho_e, \\
\mbox{curl} \left( \frac{1}{\mu_0} \mathbf{B} - \mathbf{M} \right) = \mathbf{J} + \frac{\partial}{\partial t} \left( \varepsilon_0 \mathbf{E} + \mathbf{P} \right),
\end{align}
where $\mathbf{B}$ is the magnetic induction vector, $\mathbf{E}$ is the electric field vector, $\mathbf{M}$ and $\mathbf{P}$ are, respectively, the magnetization and electric polarization of the continuum; $\mathbf{J}$ is the electric current density, $\rho_e$ is the electric charge density, $\varepsilon_0$ and $\mu_0$ are the electrical pemittivity and magnetic permeability of the vacuum respectively.  We have used the following field relations in the above equations
\begin{equation} \label{DEBM relations}
\mathbf{D} = \varepsilon_0 \mathbf{E} + \mathbf{P}, \quad \mathbf{B} = \mu_0 \left( \mathbf{H} + \mathbf{M} \right).
\end{equation}

\begin{figure}
\begin{center}
\includegraphics[scale=0.35]{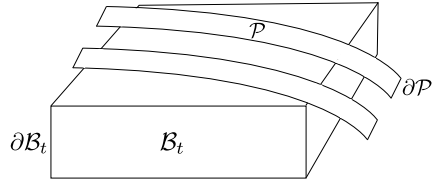}
\end{center}
\caption{A generic EMAT configuration. $\mathcal B_t$ and $\mathcal P$ are the regions occupied by the continuum material and a wire of the current-carrying coil, respectively. The corresponding boundaries are denoted by $\partial \mathcal B_t$ and $\partial \mathcal P$.}
\end{figure}

All the electromagnetic quantities above are defined in the current (Eulerian) configuration. They can also be expressed in the reference configuration (Lagrangian form) given by
\begin{eqnarray}
\mathbf{B}_l = J \mathbf{F}^{-1} \mathbf{B}, \quad \mathbf{P}_l = J \mathbf{F}^{-1} \mathbf{P}, \quad \mathbf{E}_l = \mathbf{F}^{\mathrm T} \mathbf{E}, \quad \mathbf{M}_l = \mathbf{F}^{\mathrm T} \mathbf{M}, \nonumber \\
\mathbf{J}_{\mathrm E} = J \mathbf{F}^{-1} \left( \mathbf{J} - \rho_{\mathrm e} \mathbf{v} \right), \quad \rho_{\mathrm E} = J \rho_{\mathrm e}, \label{pushback}
\end{eqnarray}
using which the Maxwell's equations can be written in Lagrangian form as
\begin{equation} \label{Lagrangian maxwell 1}
\mbox{Div}\, \mathbf{B}_l =0, \quad \mbox{Curl}\, \left( \mathbf{E}_l + \mathbf{V} \times \mathbf{B}_l \right) = -\mathbf{B}_{l,t}, \quad
\varepsilon_0 \, \mbox{Div} \left( J \mathbf{c}^{-1} \mathbf{E}_l \right) = \rho_{\mathrm E} + \mbox{Div} \, \mathbf{P}_l, 
\end{equation}
\begin{align}
\mbox{Curl}  \left(  \frac{J^{-1}}{\mu_0 } \mathbf{cB}_l - \varepsilon_0 \mathbf{V}\times (J \mathbf{c}^{-1} \mathbf{E}_l)  \right) &- \varepsilon_0 (J \mathbf{c}^{-1} \mathbf{E}_l)_{,t}  \nonumber \\
&= \mathbf{P}_{l,t} + \mbox{Curl} ( \mathbf{M}_l + \mathbf{V} \times \mathbf{P}_l ) + \mathbf{J}_{E}. \label{Lagrangian maxwell 2}
\end{align}

In the current-carrying coil of rigid material (region $\mathcal P$ in Figure 1), the governing equations for the electromagnetic fields are
\begin{equation}
\mbox{div}\, \mathbf{B} = 0, \quad \mbox{curl}\, \mathbf{E} + \frac{\partial \mathbf{B}}{\partial t} = 0, \quad \frac{1}{\mu_0 \mu_r} \mbox{curl} \, \mathbf{B} = \varepsilon_0 \varepsilon_r \frac{\partial \mathbf{E}}{\partial t} + \mathbf{J}, \quad \mbox{div}\, \mathbf{E} = 0,
\end{equation}
where the constants $\varepsilon_r$ and $\mu_r$ are the relative electric permittivity and the relative magnetic permeability of the current-carrying coil, respectively.

Outside the material, in a vacuum, the governing equations for electromagnetic fields are
\begin{equation}
\mbox{div}\, \mathbf{B}^* = 0, \quad \mbox{curl}\, \mathbf{E}^* + \frac{\partial \mathbf{B}^*}{\partial t} = \mathbf{0}, \quad \mbox{div}\, \mathbf{E}^* = 0, \quad \mbox{curl}\, \mathbf{B}^* - \varepsilon_0 \mu_0 \frac{\partial \mathbf{E}^*}{\partial t} = \mathbf{0},
\end{equation}
where we denote a physical quantity in vacuum by a superscript *.

At the boundary $\partial \mathcal B_r$, the following conditions need to be satisfied (see, for example, \cite{Saxena2012})
\begin{align}
&\mathbf{N} \times \left( \mathbf{E}_l + \mathbf{V} \times \mathbf{B}_l - \mathbf{F}^{\mathrm T} \mathbf{E}^* \right)  = \mathbf{0}, \label{Lagrangian bc 1} \\
&\mathbf{N} \cdot \left( \mathbf{B}_l - J \mathbf{F}^{-1} \mathbf{B}^* \right) = 0, \label{Lagrangian bc 2} \\
&\mathbf{N} \cdot \left\{ \varepsilon_0 J \mathbf{c}^{-1} \left( \mathbf{E}_l - \mathbf{F}^{\mathrm T} \mathbf{E}^* \right) + \mathbf{P}_l  \right\} = \sigma_{E}, \label{Lagrangian bc 3} \\
&\mathbf{N} \times \left( J^{-1}\mu_0^{-1} \mathbf{cB}_l - \mathbf{M}_l - \mathbf{V} \times (\varepsilon_0 J \mathbf{c}^{-1} \mathbf{E}_l + \mathbf{P}_l) -  \mu_0^{-1} \mathbf{F}^{\mathrm T} \mathbf{B}^* \right) = \mathbf{K}_l - \sigma_{E} \mathbf{V}_{\mathrm s}, \label{Lagrangian bc 4}
\end{align}
where $\mathbf{N}$ is the unit normal to the boundary, $\mathbf{V} = \mathbf{F}^{-1} \mathbf{v}$ is the particle velocity, $\mathbf{V}_{\mathrm s}$ is the value of $\mathbf{V}$ at the boundary, $\sigma_{ E}$ is the surface electric charge density, and $\mathbf{K}_l$ is Lagrangian description of the surface current density.
We note that $\mathbf{F}$ is not defined outside the material and the values of $\mathbf{F}$ and $J$ calculated on the boundary are used in the above equations.

The boundary conditions are simpler at $\partial \mathcal P_t$ and given by
\begin{align}
\mathbf{n} \times \left( \mathbf{E} - \mathbf{E}^* \right) = \mathbf{0}, \quad \mathbf{n} \cdot \left( \varepsilon_r \mathbf{E} - \mathbf{E}^* \right) = 0, \nonumber \\
\mathbf{n} \times \left( \frac{\mathbf{B}}{\mu_r} - \mathbf{B}^* \right) = \mathbf{K}, \quad \mathbf{n} \cdot \left( \mathbf{B} - \mathbf{B}^* \right) = 0. \label{wire boundary}
\end{align}

\subsection{Mechanical balance laws}
Balance of linear momentum, in the absence of a mechanical body force, is given by
\begin{equation}
\mbox{div}\, \boldsymbol{\tau} + \mathbf{f}_e = \rho \mathbf{a},
\end{equation}
where $\boldsymbol{\tau}$ is the Cauchy stress tensor, $\rho$ is the mass density, and  the electromagnetic body force as given in \cite{Pao1978} is 
\begin{equation}
\mathbf{f}_{e} = \rho_e \mathbf{E} + \mathbf{J} \times \mathbf{B} + (\mbox{grad} \, \mathbf{E})^{\mathrm T} \mathbf{P} + (\mbox{grad} \, \mathbf{B})^{\mathrm T} \mathbf{M} + \frac{\partial}{\partial t} (\mathbf{P} \times \mathbf{B}) + \mbox{div} [\mathbf{v} \otimes (\mathbf{P}\times \mathbf{B})].
\end{equation}

 Balance of angular momentum gives
 \begin{equation}
\boldsymbol{ \varepsilon \tau} + \mathbf{L}_{e} = \mathbf{0}, \quad \mathbf{L}_{e} = \mathbf{P} \times \mathbf{E} + (\mathbf{M}   + \mathbf{v} \times \mathbf{P} ) \times \mathbf{B},
 \end{equation}
 where $\boldsymbol{\varepsilon}$ is the third order permutation tensor with components $\varepsilon_{ijk}$ and $( \boldsymbol{\varepsilon \tau})_i  = \varepsilon_{ijk}\tau_{jk}$ and $\mathbf{L}_{e}$ is the electromagnetic body couple vector.
%  \begin{equation}
%  \mathbf{L}_{e} = \mathbf{P} \times \mathbf{E} + (\mathbf{M}   + \mathbf{v} \times \mathbf{P} ) \times \mathbf{B}.
%  \end{equation}
 The above balance equations can be written in Lagrangian form using the nominal stress tensor $\mathbf{T} = J \mathbf{F}^{-1} \boldsymbol{\tau}$ as
 \begin{equation} \label{ang and momentum balance L}
 \mbox{Div} \, \mathbf{T} + J\, \mathbf{f}_{E} = \rho_r \mathbf{a},
 \quad
 \boldsymbol{\varepsilon} (\mathbf{FT}) + J \mathbf{L}_{E} = \mathbf{0},
  \end{equation}
 where $\mathbf{f}_{E}$ and $\mathbf{L}_{E}$ are Lagrangian counterparts of the corresponding vectors and are given by
 \begin{align}
 \mathbf{f}_{E} = J^{-1} \rho_{\mathrm E} \mathbf{F}^{-\mathrm T} \mathbf{E}_l + J^{-2} (\mathbf{FJ}_l) \times (\mathbf{FB}_l) + \mathbf{F}^{- \mathrm T} \left[ \mbox{Grad}(\mathbf{F}^{-\mathrm T} \mathbf{E}_l) \right]^{\mathrm T} (J^{-1} \mathbf{FP}_l) \nonumber \\
 + \mathbf{F}^{- \mathrm T} \left[ \mbox{Grad}(J^{-1} \mathbf{FB}_l) \right]^{\mathrm T} ( \mathbf{F}^{-\mathrm T} \mathbf{M}_l) + \frac{\partial}{\partial t} \left[ J^{-2} \left(\mathbf{FP}_l \right) \times \left( \mathbf{FB}_l \right) \right] \nonumber \\
 + J^{-1} \mbox{Div} \left[ J^{-1} \mathbf{V} \otimes \left\{ (\mathbf{FP}_l) \times (\mathbf{FB}_l) \right\} \right],
 \end{align}
 \begin{equation}
 \mathbf{L}_{E} = J^{-1} \left( \mathbf{FP}_l \right)  \times \left( \mathbf{F}^{-\mathrm T} \mathbf{E}_l \right) + J^{-1}    \left( \mathbf{F}^{-\mathrm T} \mathbf{M}_{el} \right) \times \left( \mathbf{FB}_l \right).
 \end{equation}

On any part of the boundary where the traction is prescribed, the boundary condition may be given as
\begin{equation}
\mathbf{T}^{\mathrm T} \mathbf{N} = \mathbf{t}_{\mathrm A},
\end{equation}
where $\mathbf{t}_{\mathrm A}$ is the Lagrangian representation of the traction force.

\subsection{Energy balance laws and Constitutive relations}
First law of thermodynamics gives the  balance of energy as
\begin{equation} \label{FLT first}
\rho \frac{d U}{d t} = \boldsymbol{\tau} \colon \mbox{grad}\, \mathbf{v} - \mbox{div} \, \mathbf{q} + q + w_{e},
\end{equation}
where $U$ is the internal energy, $\mathbf{q}$ is the heat flux at the surface, $q$ is the volumetric heat generation, the symbol $\colon$ denotes a scalar product between two second order tensors given in   component form as $\boldsymbol{\tau}\colon \boldsymbol{\Gamma} = \tau_{ij} \Gamma_{ji}$, and $w_{e}$ is the electromagnetic power (see, for example, Pao \cite{Pao1978}) given by
\begin{equation}
w_{e} = \mathbf{J}_e \cdot \mathbf{E}_e + \rho \frac{d}{dt} \left( \frac{\mathbf{P}}{\rho} \right) \cdot \mathbf{E}_e  - \mathbf{M}_e \cdot \frac{d \mathbf{B}}{dt},
\end{equation}
where for a dynamic problem we have defined the effective field variables as
\begin{equation}
\mathbf{J}_e = \mathbf{J} - \rho_e \mathbf{v} , \quad \mathbf{E}_e = \mathbf{E} + \mathbf{v} \times \mathbf{B}, \quad \mathbf{M}_e = \mathbf{M} + \mathbf{v} \times \mathbf{P}.
\end{equation}

Let $\vartheta$ be the absolute temperature, then we can rewrite \eqref{FLT first} as
\begin{equation}
\rho \, c_p \frac{\partial \vartheta}{\partial t} =  q + w_{e} + \boldsymbol{\tau} \colon \mbox{grad}\, \mathbf{v}  - \mbox{div} \, \mathbf{q} .
\end{equation}
Here $c_p$ is the specific heat capacity.
On defining the pullback versions of the physical quantities 
\begin{equation} \label{heat pushback}
\mathbf{q}_l = J \mathbf{F}^{-1} \mathbf{q}, \quad q_l = J q, \quad w_{E} = J w_{e}, \quad \vartheta_l = J \vartheta,
\end{equation}
 the above equation can be written in  Lagrangian form as
\begin{equation} \label{heat lagrangian}
\rho_r c_p \frac{\partial }{\partial t} \left( J^{-1} \vartheta_l \right) =  \mathbf{T}\colon \mbox{Grad}(\mathbf{FV}) + q_l + w_{E} - \mbox{Div}\, \mathbf{q}_l.
\end{equation}
Here, the Lagrangian form of electromagnetic power is given as
\begin{align}
w_{E} = (\mathbf{FJ}_{el}) \cdot (\mathbf{F}^{-\mathrm T} \mathbf{E}_{el}) + \rho_r \left[ \frac{\partial}{\partial t} \left( \frac{\mathbf{FP}_l}{\rho_r} \right) + \mbox{Grad} \left( \frac{\mathbf{FP}_l}{\rho_r} \right) \mathbf{V} \right] \cdot \left( \mathbf{F}^{-\mathrm T} \mathbf{E}_{el} \right) \nonumber \\
- J \mathbf{F}^{-\mathrm T} \mathbf{M}_{el} \cdot \left[ \frac{\partial}{\partial t} \left( J^{-1} \mathbf{FB}_l \right) + \mbox{Grad} \left( J^{-1} \mathbf{FB}_l \right) \mathbf{V}  \right].
\end{align}

If $S$ is the entropy density, then the second law of thermodynamics is given by the Clausius--Duhem inequality as
\begin{equation} \label{SLT first}
\rho \frac{dS}{dt} + \mbox{div} \left( \frac{\mathbf{q}}{\vartheta} \right) - \frac{q}{\vartheta} \ge 0.
\end{equation}

Substituting equation \eqref{FLT first} into \eqref{SLT first} and following the standard Coleman-Noll procedure (see, for e.g., \cite{Coleman1963}), we arrive at the constitutive relations
\begin{equation} \label{constitutive Pl Ml}
\mathbf{T} = \frac{\partial \Phi}{\partial \mathbf{F}}, \quad \mathbf{P}_l = -  \frac{\partial \Phi}{\partial \mathbf{E}_{el}}, \quad \mathbf{M}_{el} = - \frac{\partial \Phi}{\partial \mathbf{B}_l},
\end{equation}
where $\Phi$ is the free energy per unit volume related to $U$ (see, for example, \cite{Saxena2012} for details) as
\begin{equation}
\Phi(\mathbf{F}, \mathbf{E}_{el}, \mathbf{B}_l , \vartheta_l) = \rho_r \left( U (\mathbf{F}, \mathbf{E}_e, \mathbf{B}, \vartheta) - \vartheta S - \mathbf{E}_e \cdot \frac{\mathbf{P}}{\rho} \right).
\end{equation}

We note that in the case of an incompressible material, we have an additional constraint $J=1$ and the constitutive law for stress is modified to
\begin{equation}
\mathbf{T} = \frac{\partial \Phi}{\partial \mathbf{F}} - p \mathbf{F}^{-1},
\end{equation}
$p$ being the Lagrange multiplier associated with the constraint of incompressibility.

We assume the constitutive laws governing heat flow and electric current flow to be given by the Fourier's law and the Ohm's law, respectively, as
\begin{equation}
 \mathbf{q} = -  \boldsymbol{\kappa} \, \mbox{grad}\, \vartheta, \quad \mathbf{J} = \boldsymbol{\xi} \, \mathbf{E},
\end{equation}
where $\boldsymbol{\kappa}$ and $\boldsymbol{\xi}$ are positive-definite symmetric second order tensors quantifying the thermal conductivity and the electrical conductivity, respectively. We require tensorial values for the conductivities since in general for anisotropic materials, the conductivity depends on the direction of heat flow or electric current. For isotropic materials, these assume the values $\boldsymbol{\kappa} = \kappa \mathbf{I}$ and $\boldsymbol{\xi} = \xi \mathbf{I}$. The above equations can be written in Lagrangian form as 
\begin{equation} \label{Fourier Ohm Lagrangian}
 \mathbf{q}_l = -  J \mathbf{F}^{-1} \boldsymbol{\kappa} \mathbf{F}^{- \mathrm T} \mbox{Grad}(J^{-1} \vartheta_l), \quad \mathbf{J}_l = J \mathbf{F}^{-1} \boldsymbol{\xi} \mathbf{F}^{-\mathrm T} \mathbf{E}_l.
\end{equation}

\section{Incremental equations}

On the initial motion and underlying electromagnetic fields, we consider an incremental mechanical motion $\mathbf{u}(\mathbf{x},t)$, and increments in electromagnetic fields which are denoted by a superposed dot. We emphasize here the departure from convention in using a superposed dot to denote an increment rather than a time-derivative for the sake of brevity.

The incremented forms of the Lagrangian Maxwell's equations \eqref{Lagrangian maxwell 1} and \eqref{Lagrangian maxwell 2} are given as
\begin{equation}
\mbox{Div}\, \bdot{\mathbf{B}}_l = 0, \quad  \mbox{Curl}\, \bdot{\mathbf{E}}_{el}  = - \bdot{\mathbf{B}}_{l,t},
\end{equation}
\begin{equation}
\varepsilon_0 \, \mbox{Div} \left[J (\mbox{div}\, \mathbf{u}) \mathbf{c}^{-1} \mathbf{E}_l + J \mathbf{c}^{-1} \bdot{\mathbf{E}}_l- J \mathbf{F}^{-1} (\mathbf{L+L}^{\mathrm T}) \mathbf{F}^{-\mathrm T} \mathbf{E}_l \right] = \bdot{\rho}_{\mathrm E} + \mbox{Div}\, \bdot{\mathbf{P}}_l,
\end{equation}
\begin{align}
\mu_0^{-1} \mbox{Curl} \left[J^{-1} \left\{ (\mbox{div} \, \mathbf{u}) \mathbf{cB}_l  + 2 \mathbf{F}^{\mathrm T} \mathbf{LFB}_l + \mathbf{c} \bdot{\mathbf{B}}_l  \right\} \right] - \varepsilon_0 \mbox{Curl}  \left[ \bdot{\mathbf{V}} \times (J \mathbf{c}^{-1} \mathbf{E}_l) \right. \nonumber \\
\left. + \mathbf{V} \times \left\{ J (\mbox{div} \, \mathbf{u}) \mathbf{c}^{-1} \mathbf{E}_l + J \mathbf{c}^{-1} \bdot{\mathbf{E}}_l - J \mathbf{F}^{-1} ( \mathbf{L+L}^{\mathrm T}) \mathbf{F}^{-\mathrm T} \mathbf{E}_l  \right\} \right]  \nonumber \\
- \varepsilon_0 \left[ J \left\{  J (\mbox{div} \, \mathbf{u}) \mathbf{c}^{-1} \mathbf{E}_l + J \mathbf{c}^{-1} \bdot{\mathbf{E}}_l - J \mathbf{F}^{-1} ( \mathbf{L+L}^{\mathrm T}) \mathbf{F}^{-\mathrm T} \mathbf{E}_l \right\} \right]_{,t}  \nonumber \\
= \bdot{\mathbf{P}}_{l,t} + \mbox{Curl} \, \bdot{\mathbf{M}}_{el} & + \bdot{\mathbf{J}}_{\mathrm E}.
\end{align}

The incremented Lagrangian quantities are `pushed forward' to the Eulerian configuration and denoted by a subscript `0' after `$l$' or `$E$'.
The push-forward relations for the incremented fields are given (similar to \eqref{pushback}) as
\begin{align}
\bdot{\mathbf{B}}_{l0} = J^{-1} \mathbf{F} \bdot{\mathbf{B}}_l, \quad \bdot{\mathbf{P}}_{l0} = J^{-1} \mathbf{F} \bdot{\mathbf{P}}_l,  \quad \bdot{\mathbf{M}}_{el0} = \mathbf{F}^{-\mathrm T} \bdot{\mathbf{M}}_e, \nonumber \\
 \bdot{\mathbf{E}}_{l0} = \mathbf{F}^{-\mathrm T} \bdot{\mathbf{E}}_l \quad
\bdot{\rho}_{E0} = J^{-1} \bdot{\rho}_E, \label{pushed forward}
\end{align} 
using which we can write the Eulerian form of the above incremental equations as
\begin{equation} \label{ch2 up bal 1-2}
\mbox{div}\, \bdot{\mathbf{B}}_{l0} = 0, \quad \mbox{curl}\, \bdot{\mathbf{E}}_{el0} = \left[ \boldsymbol{ \Gamma} - (\mbox{div} \, \mathbf{v}) \mathbf{I}  \right] \bdot{\mathbf{B}}_{l0} - \bdot{\mathbf{B}}_{l0,t} ,
\end{equation}
\begin{equation} \label{ch2 up bal 3}
\varepsilon_0 \mbox{div} \, \hat{\mathbf{E}}  = \bdot{\rho}_{\mathrm E 0} + \mbox{div}\, \bdot{\mathbf{P}}_{l0}, 
\end{equation}
\begin{align}
\mu_0^{-1} \mbox{curl}  \left[ \left\{ (1+ \mbox{div}\, \mathbf{u}) \mathbf{I} + 2 \mathbf{L}  \right\} \bdot{\mathbf{B}}_{l0} \right] - \varepsilon_0 \mbox{curl} \left( \mathbf{u}_{,t} \times \mathbf{E} + \mathbf{v} \times \hat{\mathbf{E}} \right) -  \varepsilon_0 \hat{\mathbf{E}}_{,t} &\nonumber \\
 = \mbox{curl} \, \bdot{\mathbf{M}}_{el0}  +  \bdot{\mathbf{P}}_{l0,t} + \left[ (\mbox{div}\, \mathbf{v}) \mathbf{I} - \boldsymbol{\Gamma} \right] \bdot{\mathbf{P}}_{l0} +  & \bdot{\mathbf{J}}_{\mathrm E0} , \label{ch2 up bal 4}
\end{align}
where
\begin{align} 
\hat{\mathbf{E}} & = \bdot{\mathbf{E}}_{l0} + (\mbox{div} \, \mathbf{u}) \mathbf{E} - (\mathbf{L+L}^{\mathrm T}) \mathbf{E},  \\
\bdot{\mathbf{E}}_{el0} & = \mathbf{F}^{-\mathrm T} \bdot{\mathbf{E}}_{el} = \bdot{\mathbf{E}}_{l0} + \mathbf{v} \times \bdot{\mathbf{B}}_{l0} + (\mathbf{u}_{,t} - \mathbf{Lv})\times \mathbf{B},
 \\
\bdot{\mathbf{J}}_{E0} &=  J^{-1} \mathbf{F} \bdot{\mathbf{J}}_E 
= \xi \hat{\mathbf{E}}  - \bdot{\rho}_{\mathrm E0} \mathbf{v} - \rho_{\mathrm e} \left( \mathbf{u}_{,t} - \mathbf{Lv} \right).
\end{align}

 The incremented momentum and angular momentum balance equations \eqref{ang and momentum balance L} are given as
\begin{equation} \label{bal 43}
\mbox{Div} \, \bdot{\mathbf{T}} + J (\mbox{div}\, \mathbf{u}) \mathbf{f}_{E} + J \bdot{\mathbf{f}}_{E} = \rho_r \bdot{\mathbf{a}},
\end{equation}
\begin{equation} \label{bal 44}
\boldsymbol{\varepsilon} (\mathbf{LFT} + \mathbf{F}\bdot{\mathbf{T}}) + J (\mbox{div}\, \mathbf{u}) \mathbf{L}_{E} + J \bdot{\mathbf{L}}_{E} = \mathbf{0}.
\end{equation}
where the increments in electromagnetic body force and moment are given by
\begin{flalign}
&\bdot{\mathbf{f}}_{E} = J^{-1} \left\{ -(\mbox{div} \, \mathbf{u}) \rho_{\mathrm E} \mathbf{F}^{- \mathrm T} \mathbf{E}_l  + \bdot{\rho}_{\mathrm E} \mathbf{F}^{- \mathrm T} \mathbf{E}_l - \rho_{\mathrm E} \mathbf{L}^{\mathrm T} \mathbf{F}^{- \mathrm T} \mathbf{E}_l + \rho_{\mathrm E} \mathbf{F}^{-\mathrm T} \bdot{\mathbf{E}}_l \right\} & \nonumber 
\end{flalign}                                                                                                                                                                                                                                                                                                                                                          \begin{equation}
- J^{-1} \mathbf{L}^{\mathrm T} \mathbf{F}^{-\mathrm T} \left[ \mbox{Grad} (\mathbf{F}^{-\mathrm T} \mathbf{E}_l) \right]^{\mathrm T} (\mathbf{FP}_l)  \nonumber \\
+ J^{-1} \mathbf{F}^{- \mathrm T} \left[ \mbox{Grad} (- \mathbf{L}^{\mathrm T} \mathbf{F}^{- \mathrm T} \mathbf{E}_l + \mathbf{F}^{- \mathrm T} \bdot{\mathbf{E}}_l ) \right]^{\mathrm T} (\mathbf{FP}_l) \nonumber 
\end{equation}                                                                                                                                                                                                                                                                                                                                                          \begin{equation}
+ J^{-1} \mathbf{F}^{-\mathrm T} \left[ \mbox{Grad} (\mathbf{F}^{-\mathrm T} \mathbf{E}_l) \right]^{\mathrm T} \left[ - (\mbox{div}\, \mathbf{u}) \mathbf{FP}_l + \mathbf{LFP}_l + \mathbf{F} \bdot{\mathbf{P}}_l \right]\nonumber 
\end{equation}                                                                                                                                                                                                                                                                                                                                                          \begin{equation}
+ 2 J^{-2} ( \mbox{div}\, \mathbf{u}) (\mathbf{FJ}_l) \times (\mathbf{FB}_l) + J^{-2} \left( \mathbf{LFJ}_l + \mathbf{F} \bdot{\mathbf{J}}_l \right) \times (\mathbf{FB}_l) \nonumber 
\end{equation}                                                                                                                                                                                                                                                                                                                                                          \begin{equation}
 + J^{-2} (\mathbf{FJ}_l) \times (\mathbf{LFB}_l + \mathbf{F} \bdot{\mathbf{B}}_l ) 
- \mathbf{L}^{\mathrm T} \mathbf{F}^{-\mathrm T} \left[ \mbox{Grad} \left( J^{-1} \mathbf{FB}_l \right) \right]^{\mathrm T} (\mathbf{F}^{-\mathrm T} \mathbf{M}_l) \nonumber 
\end{equation}                                                                                                                                                                                                                                                                                                                                                          \begin{equation}
+ \mathbf{F}^{-\mathrm T} \left[ \mbox{Grad} \left( - J^{-1} (\mbox{div}\, \mathbf{u}) \mathbf{FB}_l + J^{-1} \mathbf{LFB}_l + J^{-1} \mathbf{F} \bdot{\mathbf{B}}_l  \right) \right]^{\mathrm T} (\mathbf{F}^{-\mathrm T} \mathbf{M}_l ) \nonumber 
\end{equation}                                                                                                                                                                                                                                                                                                                                                          \begin{equation}
+ \mathbf{F}^{- \mathrm T} \left[ \mbox{Grad} (J^{-1} \mathbf{FB}_l) \right]^{\mathrm T} \left( - \mathbf{L}^{\mathrm T} \mathbf{F}^{- \mathrm T} \mathbf{M}_l + \mathbf{F}^{- \mathrm T} \bdot{\mathbf{M}}_l \right) \nonumber 
\end{equation}                                                                                                                                                                                                                                                                                                                                                          \begin{equation}
+ \frac{\partial}{\partial t} \left[ 2 J^{-2} (\mbox{div} \, \mathbf{u}) \left( \mathbf{FP}_l \right) \times \left( \mathbf{FB}_l \right) + J^{-2} \left( \mathbf{LFP}_l + \mathbf{F} \bdot{\mathbf{P}}_l \right) \times \left( \mathbf{FB}_l \right) \right.  \nonumber 
\end{equation}                                                                                                                                                                                                                                                                                                                                                          \begin{equation}
\left. + J^{-2} \left( \mathbf{FP}_l \right) \times \left( \mathbf{LFB}_l + \mathbf{F} \bdot{\mathbf{B}}_l  \right) \right] 
- J^{-1} (\mbox{div}\, \mathbf{u}) \, \mbox{Div} \left[ J^{-1}  \mathbf{V} \otimes \left\{ (\mathbf{FP}_l \times (\mathbf{FB}_l) \right\} \right] \nonumber
\end{equation}                                                                                                                                                                                                                                                                                                                                                          \begin{equation}
 + J^{-1} \mbox{Div} \left[ J^{-1} \bdot{\mathbf{V}} \otimes \left\{ (\mathbf{FP}_l) \times (\mathbf{FB}_l) \right\} 
 - J^{-1} (\mbox{div}\, \mathbf{u}) \mathbf{V} \otimes \left\{ \left(\mathbf{FP}_l \right) \times \left( \mathbf{FB}_l  \right) \right\} \right. \nonumber 
\end{equation}                                                                                                                                                                                                                                                                                                                                                          \begin{equation}
 \left. + J^{-1} \mathbf{V} \otimes \left\{ \left(\mathbf{LFP}_l + \mathbf{F} \bdot{\mathbf{P}}_l \right) \times \left( \mathbf{FB}_l \right) 
  + \left( \mathbf{FP}_l \right) \times \left( \mathbf{LFB}_l + \mathbf{F} \bdot{\mathbf{B}}_l \right)  \right\} \right],
\end{equation}
\begin{flalign}
&\bdot{\mathbf{L}}_{E} = - J^{-1} (\mbox{div}\, \mathbf{u}) \left\{ \left( \mathbf{FP}_l \right) \times \left( \mathbf{F}^{-\mathrm T} \mathbf{E}_l  \right)    + \left( \mathbf{F}^{- \mathrm T} \mathbf{M}_{el} \right) \times \left( \mathbf{FB}_l \right) \right\} & \nonumber 
\end{flalign}
\begin{equation}
+ J^{-1} \left( \mathbf{LFP}_l + \mathbf{F} \bdot{\mathbf{P}}_l \right) \times \left(\mathbf{F}^{-\mathrm T} \mathbf{E}_l \right) + J^{-1} \left(\mathbf{FP}_l \right)\times \left( - \mathbf{L}^{\mathrm T} \mathbf{F}^{-\mathrm T} \mathbf{E}_l + \mathbf{F}^{-\mathrm T} \bdot{\mathbf{E}}_l \right)  \nonumber 
\end{equation}
\begin{equation}
 + J^{-1} \left( \mathbf{F}^{-\mathrm T} \mathbf{M}_{el} \right) \times \left( \mathbf{LFB}_l + \mathbf{F} \bdot{\mathbf{B}}_l \right) + J^{-1} \left( - \mathbf{L}^{\mathrm T} \mathbf{F}^{-\mathrm T} \mathbf{M}_{el} + \mathbf{F}^{-\mathrm T} \bdot{\mathbf{M}}_{el} \right) \times \left(\mathbf{FB}_l \right) .
\end{equation}

 When updated to Eulerian form, the balance equations \eqref{bal 43} and \eqref{bal 44} become
\begin{equation} \label{ch2 up bal 5}
\mbox{div}\, \bdot{\mathbf{T}}_0 + (\mbox{div}\, \mathbf{u}) \mathbf{f}_{e} + \bdot{\mathbf{f}}_{E0} = \rho \mathbf{u}_{,tt},
\end{equation}
\begin{equation} \label{ch2 up bal 6}
\boldsymbol{\varepsilon} (\mathbf{L} \boldsymbol{\tau} + \bdot{\mathbf{T}}_0) + (\mbox{div}\, \mathbf{u}) \mathbf{L}_{e} + \bdot{\mathbf{L}}_{E0} = \mathbf{0},
\end{equation}
where $\bdot{\mathbf{f}}_{E0}$ and $\bdot{\mathbf{L}}_{E0}$ are the push-forward forms of the incremental body force and moment, respectively, and are given by
\begin{flalign}
& \bdot{\mathbf{f}}_{E0} =  - (\mbox{div}\, \mathbf{u}) \rho_{\mathrm e} \mathbf{E} + \bdot{\rho}_{E0} \mathbf{E} - \rho_{\mathrm e} \mathbf{L}^{\mathrm T} \mathbf{E} + \rho_{\mathrm e} \bdot{\mathbf{E}}_{l0} - \mathbf{L}^{\mathrm T} \left( \mbox{grad}\, \mathbf{E}\right)^{\mathrm T} \mathbf{P} & \nonumber 
\end{flalign}
\begin{equation}
+ \left[ \mbox{grad} \left( - \mathbf{L}^{\mathrm T} \mathbf{E} + \bdot{\mathbf{E}}_{l0} \right) \right]^{\mathrm T} \mathbf{P} + \left( \mbox{grad}\, \mathbf{E} \right)^{\mathrm T} \left[ - (\mbox{div}\, \mathbf{u}) \mathbf{P} + \mathbf{LP} + \bdot{\mathbf{P}}_{l0}  \right] \nonumber 
\end{equation}
\begin{equation}
-2 (\mbox{div}\, \mathbf{u})\,  \mathbf{J} \times \mathbf{B} + \left( \mathbf{LJ} + \bdot{\mathbf{J}}_{l0} \right) \times \mathbf{B} + \mathbf{J} \times \left( \mathbf{LB} + \bdot{\mathbf{B}}_{l0} \right) - \mathbf{L}^{\mathrm T} \left( \mbox{grad}\, \mathbf{B} \right)^{\mathrm T} \mathbf{M} \nonumber 
\end{equation}
\begin{equation}
+ \left[ \mbox{grad} \left( - (\mbox{div}\, \mathbf{u}) \mathbf{B} + \mathbf{LB} + \bdot{\mathbf{B}}_{l0} \right) \right]^{\mathrm T} \mathbf{M} + \left( \mbox{grad}\, \mathbf{B} \right)^{\mathrm T} \left(  - \mathbf{L}^{\mathrm T} \mathbf{M} + \bdot{\mathbf{M}}_{l0} \right) \nonumber
\end{equation}
\begin{equation}
+ \frac{\partial}{\partial t} \left[ 2 (\mbox{div}\, \mathbf{u}) \mathbf{P}\times \mathbf{B} + \left(\mathbf{LP} + \bdot{\mathbf{P}}_{l0} \right) \times \mathbf{B} + \mathbf{P} \times \left( \mathbf{LB} + \bdot{\mathbf{B}}_{l0} \right) \right] \nonumber
\end{equation}
\begin{equation}
-(\mbox{div}\, \mathbf{u}) \, \mbox{div} \left[ \mathbf{v} \otimes (\mathbf{P}\times \mathbf{B}) \right] + \mbox{div} \left[  \left( \bdot{\mathbf{v}} - \mathbf{Lv} \right) \otimes (\mathbf{P} \times \mathbf{B}) - (\mbox{div}\, \mathbf{u})  \mathbf{v} \otimes (\mathbf{P} \times \mathbf{B})  \right. \nonumber 
\end{equation}
\begin{equation}
\left. + \mathbf{v} \otimes \left\{ \left(\mathbf{LP} + \bdot{\mathbf{P}}_{l0} \right) \times \mathbf{B} \right\} + \mathbf{P} \times \left( \mathbf{LB} + \bdot{\mathbf{B}}_{l0} \right) \right],
\end{equation}
\begin{flalign}
& \bdot{\mathbf{L}}_{E0} = - (\mbox{div}\, \mathbf{u}) \left( \mathbf{P} \times \mathbf{E} + \mathbf{M}_e \times \mathbf{B} \right)
+ \left( \mathbf{LP} + \bdot{\mathbf{P}}_{l0} \right) \times \mathbf{E} + \mathbf{P} \times \left( - \mathbf{L}^{\mathrm T} \mathbf{E} + \bdot{\mathbf{E}}_{l0} \right) & \nonumber 
\end{flalign}
\begin{equation}
+ \mathbf{M}_e \times \left( \mathbf{LB} + \bdot{\mathbf{B}}_{l0}  \right) + \left( - \mathbf{L}^{\mathrm T} \mathbf{M}_e + \bdot{\mathbf{M}}_{el0} \right) \times \mathbf{B}.
\end{equation}

The heat equation \eqref{heat lagrangian} can be incremented to give
\begin{align}
\mbox{Div}\, \bdot{\mathbf{q}}_l + \rho_r c_p \frac{\partial}{\partial t} \left[ J^{-1} \bdot{\vartheta}_l - J^{-1} (\mbox{div}\, \mathbf{u}) \vartheta_l \right] =  \bdot{q}_l + \bdot{w}_{E} + \bdot{\mathbf{T}} \colon \mbox{Grad} (\mathbf{FV})  \nonumber \\
+ \mathbf{T} \colon \mbox{Grad} (\mathbf{LFV} + \mathbf{F} \bdot{\mathbf{V}}),
\end{align}
which when updated to the Eulerian configuration becomes
\begin{equation} \label{heat updated}
\mbox{div} \, \bdot{\mathbf{q}}_{l0} + \rho c_p \frac{\partial}{\partial t} \left[ \bdot{\vartheta}_{l0} - (\mbox{div}\, \mathbf{u}) \vartheta \right] = \bdot{q}_{l0} + \bdot{w}_{E0} + \bdot{\mathbf{T}}_0 \colon \mbox{grad}\, \mathbf{v} + \boldsymbol{\tau} \colon \mbox{grad}\, \bdot{\mathbf{v}}.
\end{equation}
We have used the push-forward relations  $ \bdot{\mathbf{q}}_{l0} = J^{-1} \mathbf{F} \bdot{\mathbf{q}}_l, \bdot{\vartheta}_{l0} = J^{-1} \bdot{\vartheta}_l , \bdot{w}_{E0} = J^{-1} \bdot{w}_{E}$, and $\bdot{q}_{l0} = J^{-1} \bdot{q}_l$   to effect the above transformation. The increment in the electromagnetic power is given by
\begin{flalign}
&\bdot{w}_{E} = \left( \mathbf{LFJ}_{el} + \mathbf{F} \bdot{\mathbf{J}}_{el} \right) \cdot \left( \mathbf{F}^{-\mathrm T} \mathbf{E}_{el} \right) 
+ \left( \mathbf{FJ}_{el} \right) \cdot \left( - \mathbf{L}^{\mathrm T} \mathbf{F}^{-\mathrm T} \mathbf{E}_{el} + \mathbf{F}^{-\mathrm T} \cdot \bdot{\mathbf{E}}_{el} \right) & \nonumber
\end{flalign}
\begin{equation}
+ \rho_r \left[  \frac{\partial}{\partial t} \left( \frac{\mathbf{LFP}_l + \mathbf{F} \bdot{\mathbf{P}}_l}{\rho_r} \right) + \mbox{Grad} \left( \frac{\mathbf{LFP}_l + \mathbf{F} \bdot{\mathbf{P}}_l}{\rho_r} \right) \mathbf{V} + \mbox{Grad} \left( \frac{\mathbf{FP}_l}{\rho_r} \right) \bdot{\mathbf{V}}\right] \nonumber  
\end{equation}
\begin{equation}
\cdot \left( \mathbf{F}^{-\mathrm T} \mathbf{E}_{el} \right)  + \rho_r \left[ \frac{\partial}{\partial t} \left( \frac{\mathbf{FP}_l}{\rho_r} \right) + \mbox{Grad} \left( \frac{\mathbf{FP}_l}{\rho_r} \right) \mathbf{V}\right] 
\cdot \left( - \mathbf{L}^{\mathrm T} \mathbf{F}^{-\mathrm T} \mathbf{E}_{el} + \mathbf{F}^{-\mathrm T} \cdot \bdot{\mathbf{E}}_{el} \right) \nonumber 
\end{equation}
\begin{equation}
- J\left( (\mbox{div}\, \mathbf{u}) \mathbf{F}^{-\mathrm T} \mathbf{M}_{el} - \mathbf{L}^{\mathrm T} \mathbf{F}^{-\mathrm T} \mathbf{M}_{el} + \mathbf{F}^{-\mathrm T} \bdot{\mathbf{M}}_{el}  \right) \cdot \left[ \frac{\partial}{\partial t} \left( J^{-1} \mathbf{FB}_l \right) \right. \nonumber 
\end{equation}
\begin{equation}
\left. + \mbox{Grad} \left( J^{-1} \mathbf{FB}_l \right) \mathbf{V} \right] - J \mathbf{F}^{-\mathrm T} \mathbf{M}_{el} \cdot \left[ \frac{\partial}{\partial t} J^{-1} \left( -(\mbox{div}\, \mathbf{u}) \mathbf{FB}_l + \mathbf{LFB}_l + \mathbf{F} \bdot{\mathbf{B}}_l  \right) \right. \nonumber 
\end{equation}
\begin{equation}
\left. + \mbox{Grad}\,  J^{-1} \left\{  -(\mbox{div}\, \mathbf{u}) \mathbf{FB}_l + \mathbf{LFB}_l + \mathbf{F} \bdot{\mathbf{B}}_l  \right\} \mathbf{V} + \mbox{Grad} \left( J^{-1} \mathbf{FB}_l \right) \bdot{\mathbf{V}} \right].
\end{equation}
This is given in Eulerian form as
\begin{flalign}
&\bdot{w}_{E0} = \left( \mathbf{LJ}_e + \bdot{\mathbf{J}}_{el0} \right) \cdot \mathbf{E}_e + \mathbf{J}_e \cdot \left( - \mathbf{L}^{\mathrm T} \mathbf{E}_e + \bdot{\mathbf{E}}_{el0} \right) &\nonumber 
\end{flalign}
\begin{equation}
\rho \left[ \frac{d}{dt} \left( \frac{\mathbf{LP} + \bdot{\mathbf{P}}_{l0} }{\rho} \right) + \mbox{grad} \left( \frac{\mathbf{P}}{\rho} \right) \left( \bdot{\mathbf{v}} - \mathbf{Lv} \right) \right] \cdot \mathbf{E}_e + \rho \frac{d}{dt} \left( \frac{\mathbf{P}}{\rho} \right) \cdot \left( -\mathbf{L}^{\mathrm T} \mathbf{E}_e + \bdot{\mathbf{E}}_{el0} \right) \nonumber 
\end{equation}
\begin{equation}
- \left\{ (\mbox{div}\, \mathbf{u}) \mathbf{M}_e - \mathbf{L}^{\mathrm T} \mathbf{M}_e + \bdot{\mathbf{M}}_{el0} \right\} \cdot \frac{d \mathbf{B}}{dt} - \mathbf{M}_e \cdot \left[ \frac{d}{dt} \left\{ -(\mbox{div}\, \mathbf{u}) \mathbf{B} + \mathbf{LB} + \bdot{\mathbf{B}}_{l0} \right\} \right. \nonumber
\end{equation}
\begin{flalign}
&&\left. + \left( \mbox{grad}\, \mathbf{B} \right) \left( \bdot{\mathbf{v}} - \mathbf{Lv} \right) \right].&
\end{flalign}

\subsection{Incremental constitutive equations and Moduli tensors}

On incrementing the constitutive equations \eqref{constitutive Pl Ml}, we obtain
\begin{equation} \label{emat incremented const 1, 2}
\bdot{\mathbf{T}} =   \boldsymbol{\mathcal A} \bdot{\mathbf{F}} + \boldsymbol{\mathcal B} \bdot{\mathbf{E}}_{el} + \boldsymbol{\mathcal C} \bdot{\mathbf{B}}_l + \boldsymbol{\mathcal D} \bdot{\vartheta}_l  , \quad \bdot{\mathbf{P}}_{l} = - \left( \boldsymbol{\mathcal F} \bdot{\mathbf{F}} + \boldsymbol{\mathcal G} \bdot{\mathbf{E}}_{el} + \boldsymbol{\mathcal H} \bdot{\mathbf{B}}_l + \boldsymbol{\mathcal I} \bdot{\vartheta}_l \right) ,
\end{equation}
and
\begin{equation} \label{emat incremented const 3}
\bdot{\mathbf{M}}_{el} = - \left(  \boldsymbol{\mathcal K} \bdot{\mathbf{F}} + \boldsymbol{\mathcal L} \bdot{\mathbf{E}}_{el} + \boldsymbol{\mathcal M} \bdot{\mathbf{B}}_l + \boldsymbol{\mathcal N} \bdot{\vartheta}_l  \right) ,
\end{equation}
where we have defined the moduli tensors as
\begin{align}
\boldsymbol{\mathcal A} = \frac{\partial^2 \Phi}{\partial \mathbf{F} \partial \mathbf{F}}, \quad \boldsymbol{\mathcal B} = \frac{\partial^2 \Phi}{\partial \mathbf{E}_{el} \partial \mathbf{F}}, \quad \boldsymbol{\mathcal C} = \frac{\partial^2 \Phi}{\partial \mathbf{B}_l \partial \mathbf{F}}, \quad \boldsymbol{\mathcal D} = \frac{\partial^2 \Phi}{\partial \vartheta_l \partial \mathbf{F}},  
\nonumber \\
\boldsymbol{\mathcal F} = \frac{\partial^2 \Phi}{\partial \mathbf{F} \partial \mathbf{E}_{el}}, \quad \boldsymbol{\mathcal G} = \frac{\partial^2 \Phi}{\partial \mathbf{E}_{el} \partial \mathbf{E}_{el}}, \quad \boldsymbol{\mathcal H} = \frac{\partial^2 \Phi}{\partial \mathbf{B}_l \partial \mathbf{E}_{el}}, \quad \boldsymbol{\mathcal I} = \frac{\partial^2 \Phi}{\partial \vartheta_l \partial \mathbf{E}_{el}},
\nonumber \\
\boldsymbol{\mathcal K} = \frac{\partial^2 \Phi}{\partial \mathbf{F} \partial \mathbf{B}_l}, \quad \boldsymbol{\mathcal L} = \frac{\partial^2 \Phi}{\partial \mathbf{E}_{el} \partial \mathbf{B}_l}, \quad \boldsymbol{\mathcal M} = \frac{\partial^2 \Phi}{\partial \mathbf{B}_l \partial \mathbf{B}_l}, \quad \boldsymbol{\mathcal N} = \frac{\partial^2 \Phi}{\partial \vartheta_l \partial \mathbf{B}_l}. 
\end{align}

Here $\boldsymbol{\mathcal A}$ is a fourth-order tensor, $\boldsymbol{\mathcal B}, \boldsymbol{\mathcal C}, \boldsymbol{\mathcal F}, \boldsymbol{\mathcal K}$ are third-order tensors, $\boldsymbol{\mathcal G}, \boldsymbol{\mathcal H}, \boldsymbol{\mathcal L}, \boldsymbol{\mathcal M}$ are second-order tensors, and $\boldsymbol{\mathcal D}, \boldsymbol{\mathcal I}, \boldsymbol{\mathcal N}$ are first-order tensors (or vectors).
To put them in perspective, for a problem involving simpler quasimagnetostatic case as presented in \cite{Dorfmann2004} and \cite{Saxena2011}, only $\boldsymbol{\mathcal A}, \boldsymbol{\mathcal C}$, and $\boldsymbol{\mathcal M}$ are required.
 Products in \eqref{emat incremented const 1, 2} and \eqref{emat incremented const 3} are defined in component form as
\begin{align}
& (\boldsymbol{\mathcal A} \bdot{\mathbf{F}})_{\alpha i} = {\mathcal A}_{\alpha i  \beta j } \bdot{{F}}_{j \beta}, \quad (\boldsymbol{\mathcal B} \bdot{\mathbf{E}}_{el})_{\alpha i} = {\mathcal B}_{\alpha i | \beta} \bdot{{E}}_{el \beta}, \quad (\boldsymbol{\mathcal C} \bdot{\mathbf{B}}_l)_{\alpha i} = {\mathcal C}_{\alpha i| \beta} \bdot{{B}}_{l \beta}, 
  \nonumber \\
& (\boldsymbol{\mathcal F} \bdot{\mathbf{F}})_i = {\mathcal F}_{i| \alpha j} \bdot{{F}}_{j \alpha}, \quad (\boldsymbol{\mathcal G} \bdot{\mathbf{E}}_{el})_\alpha = {\mathcal G}_{\alpha \beta} \bdot{{E}}_{el \beta}, \quad (\boldsymbol{\mathcal H} \bdot{\mathbf{B}}_l)_\alpha = {\mathcal H}_{\alpha \beta} \bdot{{B}}_{l \beta},
 \nonumber \\
& (\boldsymbol{\mathcal K} \bdot{\mathbf{F}})_i = {\mathcal K}_{i| \alpha j} \bdot{{F}}_{j \alpha}, \quad (\boldsymbol{\mathcal L} \bdot{\mathbf{E}}_{el})_\alpha = {\mathcal L}_{\alpha \beta} \bdot{{E}}_{el \beta}, \quad (\boldsymbol{\mathcal M} \bdot{\mathbf{B}}_l)_\alpha = {\mathcal M}_{\alpha \beta} \bdot{{B}}_{l \beta}, 
\end{align}
and the following relations hold
\begin{equation}
\boldsymbol{\mathcal K} = \boldsymbol{\mathcal C}^{\mathrm T}, \quad \boldsymbol{\mathcal F} = \boldsymbol{\mathcal B}^{\mathrm T}, \quad \boldsymbol{\mathcal L} = \boldsymbol{\mathcal H}^{\mathrm T}.
\end{equation}

On updating (push-forward to Eulerian configuration) the incremented constitutive equations \eqref{emat incremented const 1, 2} and \eqref{emat incremented const 3}, we obtain
\begin{equation} \label{emat up incremented const 1}
\bdot{\mathbf{T}}_0 =   \boldsymbol{\mathcal A}_0 \mathbf{L} + \boldsymbol{\mathcal B}_0 \bdot{\mathbf{E}}_{el0} + \boldsymbol{\mathcal C}_0 \bdot{\mathbf{B}}_{l0} + \boldsymbol{\mathcal D}_0 \bdot{\vartheta}_{l0}, 
\end{equation}
\begin{equation} \label{emat up incremented const 2}
\bdot{\mathbf{P}}_{l0} = - \left( \boldsymbol{\mathcal B}_0^{\mathrm T} \mathbf{L} + \boldsymbol{\mathcal G}_0 \bdot{\mathbf{E}}_{el0} + \boldsymbol{\mathcal H}_0 \bdot{\mathbf{B}}_{l0} + \boldsymbol{\mathcal I}_0 \bdot{\vartheta}_{l0} \right),
\end{equation}
and
\begin{equation} \label{emat up incremented const 3}
\bdot{\mathbf{M}}_{el0} = - \left( \boldsymbol{\mathcal C}_0^{\mathrm T} \mathbf{L} + \boldsymbol{\mathcal H}_0^{\mathrm T} \bdot{\mathbf{E}}_{el0} + \boldsymbol{\mathcal M}_0 \bdot{\mathbf{B}}_{l0} + \boldsymbol{\mathcal N}_0 \bdot{\vartheta}_{l0} \right), 
\end{equation}
with $\bdot{\vartheta}_{l0} = J \bdot{\vartheta}_l$ and the updated moduli tensors defined in component form as
\begin{align}
\mathcal A_{0piqj} = J^{-1} F_{p \alpha} F_{q \beta} \mathcal A_{\alpha i \beta j}, \quad \mathcal B_{0ij|k} = J^{-1} F_{i \alpha} F_{k \beta} \mathcal B_{\alpha j| \beta}, \quad \mathcal C_{0ij|k} =  F_{i \alpha} F_{\beta k}^{-1} \mathcal C_{\alpha j | \beta}, \nonumber \\
\mathcal D_{0ij} = F_{ik} \mathcal D_{kj}  , \quad \mathcal G_{0ij} = J^{-1} F_{i \alpha} F_{j \beta} \mathcal G_{\alpha \beta}, \quad \mathcal H_{0ij} = F_{i \alpha} F_{\beta j}^{-1} \mathcal H_{\alpha \beta}, \nonumber \\
 \mathcal I_{0i} = F_{ik} \mathcal I_{k}, \quad \mathcal M_{0ij} = J F_{\alpha i}^{-1} F_{\beta j}^{-1} \mathcal M_{\alpha \beta},  \quad \mathcal N_{0i} = J^{-1} F^{-1}_{ki} \mathcal N_{k}.
\end{align}

Incrementing and updating the constitutive equations \eqref{Fourier Ohm Lagrangian}, we get
\begin{equation} \label{66}
\bdot{\mathbf{q}}_{l0} = -(\mbox{div}\, \mathbf{u}) \boldsymbol{\kappa} \, \mbox{grad}\, \vartheta + 2 \mathbf{L} \boldsymbol{\kappa} \, \mbox{grad}\, \vartheta - \boldsymbol{\kappa} \, \mbox{grad} \left[ \bdot{\vartheta}_{l0} - (\mbox{div}\, \mathbf{u}) \vartheta \right],
\end{equation}
and
\begin{equation} \label{67}
\bdot{\mathbf{J}}_{l0} = \boldsymbol{\xi} \left[ \left\{ (\mbox{div}\, \mathbf{u}) \mathbf{I} - \mathbf{L} - \mathbf{L}^{\mathrm T} \right\} \mathbf{E} + \bdot{\mathbf{E}}_{l0} \right].
\end{equation}

On substituting the incremented updated constitutive equations \eqref{emat up incremented const 1}, \eqref{emat up incremented const 2}, \eqref{emat up incremented const 3}, \eqref{66}, and \eqref{67} into the incremented updated balance equations \eqref{ch2 up bal 3}, \eqref{ch2 up bal 4}, \eqref{ch2 up bal 5}, \eqref{ch2 up bal 6}, and \eqref{heat updated}, we obtain
\begin{align} \label{ch6 general bal 3}
\varepsilon_0 \, \mbox{div} \left[ \bdot{\mathbf{E}}_{l0} + \left\{ (\mbox{div} \, \mathbf{u}) \mathbf{I} - (\mathbf{L+L}^{\mathrm T}) \right\} \mathbf{E} \right]&  = \bdot{\rho}_{E0} \nonumber \\ 
- \mbox{div} & \left( \boldsymbol{\mathcal B}_0^{\mathrm T} \mathbf{L} + \boldsymbol{\mathcal G}_0 \bdot{\mathbf{E}}_{el0} + \boldsymbol{\mathcal H}_0 \bdot{\mathbf{B}}_{l0} + \boldsymbol{\mathcal I}_0 \bdot{\vartheta}_{l0} \right),
\end{align}
\begin{align} \label{ch6 general bal 4}
\mu_0^{-1} \mbox{curl}  \left[ \left\{ (1+ \mbox{div}\, \mathbf{u}) \mathbf{I} + 2 \mathbf{L}  \right\} \bdot{\mathbf{B}}_{l0} \right] - \varepsilon_0 \mbox{curl} \left( \mathbf{u}_{,t} \times \mathbf{E} + \mathbf{v} \times \hat{\mathbf{E}} \right) - \varepsilon_0 \hat{\mathbf{E}}_{,t} \nonumber \\
= - \mbox{curl} \left( \boldsymbol{\mathcal C}_0^{\mathrm T} \mathbf{L} + \boldsymbol{\mathcal H}_0^{\mathrm T} \bdot{\mathbf{E}}_{el0} + \boldsymbol{\mathcal M}_0 \bdot{\mathbf{B}}_{l0} + \boldsymbol{\mathcal N}_0 \bdot{\vartheta}_{l0} \right)   + \bdot{\mathbf{J}}_{\mathrm E 0}  \nonumber \\
- \left[ (\mbox{div}\, \mathbf{u}) \mathbf{I} - \boldsymbol{\Gamma} \right] \left( \boldsymbol{\mathcal B}_0^{\mathrm T} \mathbf{L} + \boldsymbol{\mathcal G}_0 \bdot{\mathbf{E}}_{el0} + \boldsymbol{\mathcal H}_0 \bdot{\mathbf{B}}_{l0} + \boldsymbol{\mathcal I}_0 \bdot{\vartheta}_{l0}  \right) \nonumber \\
 - \left(  \boldsymbol{\mathcal B}_0^{\mathrm T} \mathbf{L} + \boldsymbol{\mathcal G}_0 \bdot{\mathbf{E}}_{el0} + \boldsymbol{\mathcal H}_0 \bdot{\mathbf{B}}_{l0} + \boldsymbol{\mathcal I}_0 \bdot{\vartheta}_{l0}  \right)_{,t},
\end{align}
\begin{equation}  \label{ch6 general bal 5}
\mbox{div}\, \left( \boldsymbol{\mathcal A}_0 \mathbf{L} + \boldsymbol{\mathcal B}_0 \bdot{\mathbf{E}}_{el0} + \boldsymbol{\mathcal C}_0 \bdot{\mathbf{B}}_{l0} + \boldsymbol{\mathcal D}_0 \bdot{\vartheta}_{l0}  \right)  + (\mbox{div}\, \mathbf{u}) \mathbf{f}_{e} + \bdot{\mathbf{f}}_{E0} = \rho \mathbf{u}_{,tt},
\end{equation}
\begin{equation}  \label{ch6 general bal 6}
\boldsymbol{\varepsilon} \left(\mathbf{L} \boldsymbol{\tau} + \boldsymbol{\mathcal A}_0 \mathbf{L} + \boldsymbol{\mathcal B}_0 \bdot{\mathbf{E}}_{el0} + \boldsymbol{\mathcal C}_0 \bdot{\mathbf{B}}_{l0} + \boldsymbol{\mathcal D}_0 \bdot{\vartheta}_{l0} \right) + (\mbox{div}\, \mathbf{u}) \mathbf{L}_{e} + \bdot{\mathbf{L}}_{E0} = \mathbf{0},
\end{equation}
\begin{align} \label{ch6 general bal 7}
\mbox{div} \left[  -(\mbox{div}\, \mathbf{u}) \boldsymbol{\kappa} \, \mbox{grad}\, \vartheta + 2 \mathbf{L} \boldsymbol{\kappa} \, \mbox{grad}\, \vartheta - \boldsymbol{\kappa} \, \mbox{grad} \left( \bdot{\vartheta}_{l0} - (\mbox{div}\, \mathbf{u}) \vartheta \right) \right]  \nonumber \\
= -\rho c_p \frac{\partial}{\partial t} \left[ \bdot{\vartheta}_{l0} - (\mbox{div}\, \mathbf{u}) \vartheta \right]  + \bdot{q}_{l0} + \bdot{w}_{E0}   + \boldsymbol{\tau} \colon \mbox{grad}\, \mathbf{u}_{,t} \nonumber \\
+ \left( \boldsymbol{\mathcal A}_0 \mathbf{L} + \boldsymbol{\mathcal B}_0 \bdot{\mathbf{E}}_{el0} + \boldsymbol{\mathcal C}_0 \bdot{\mathbf{B}}_{l0} + \boldsymbol{\mathcal D}_0 \bdot{\vartheta}_{l0}   \right) \colon \mbox{grad}\, \mathbf{v},
\end{align}
along with
\eqref{ch2 up bal 1-2}.
The governing equations of the incremental fields in the region $\mathcal P$ are
\begin{equation}
\mbox{div}\, \bdot{\mathbf{B}} = 0,  \quad \varepsilon_r \, \mbox{div} \, \bdot{\mathbf{E}} = 0, \quad
\mbox{curl}\, \bdot{\mathbf{E}} = - \frac{\partial \bdot{\mathbf{B}}}{\partial t},  \quad \frac{1}{\mu_0 \mu_r} \mbox{curl}\, \bdot{\mathbf{B}} = \bdot{ \mathbf{J}} + \varepsilon_0 \varepsilon_r \frac{\partial \bdot{\mathbf{E}}}{\partial t}, \label{wire increm maxwell}
\end{equation}
while in vacuum, we have
\begin{equation} \label{vacuum increm maxwell}
\mbox{div}\, \bdot{\mathbf{B}}^* = 0,  \quad  \mbox{div} \, \bdot{\mathbf{E}}^* = 0, \quad
\mbox{curl}\, \bdot{\mathbf{E}}^* = - \frac{\partial \bdot{\mathbf{B}}^*}{\partial t}, \quad  \mbox{curl}\, \bdot{\mathbf{B}}^* =   \varepsilon_0 \mu_0 \frac{\partial \bdot{\mathbf{E}}^*}{\partial t}.
\end{equation}

\subsection{Incremental boundary conditions}
At the boundary $\partial \mathcal B_r$ of the continuum and vacuum, the incremental form of the boundary conditions \eqref{Lagrangian bc 1}--\eqref{Lagrangian bc 4} is given as
\begin{equation}
 \mathbf{N} \times \left( \bdot{\mathbf{E}}_l + \bdot{\mathbf{V}}\times \mathbf{B}_l + \mathbf{V} \times \bdot{\mathbf{B}}_l - \mathbf{F}^{\mathrm T} \mathbf{LE}^* - \mathbf{F}^{\mathrm T} \bdot{\mathbf{E}}^* \right) = \mathbf{0},
 \end{equation}
 \begin{equation}
 \mathbf{N} \cdot \left( \bdot{\mathbf{B}}_l - J (\mbox{div}\, \mathbf{u}) \mathbf{F}^{-1} \mathbf{B}^* - J  \mathbf{F}^{-1} \mathbf{L}^{-1} \mathbf{B}^* - J \mathbf{F}^{-1} \bdot{\mathbf{B}}^* \right) = 0,
 \end{equation}
 \begin{align}
 \mathbf{N} \cdot \left\{ \varepsilon_0 J \mathbf{c}^{-1} \left( \bdot{\mathbf{E}}_l - \mathbf{F}^{\mathrm T} \mathbf{L}^{\mathrm T} \mathbf{E}^* - \mathbf{F}^{\mathrm T} \bdot{\mathbf{E}}^* \right) + \varepsilon_0 J (\mbox{div}\, \mathbf{u}) \mathbf{c}^{-1} \left( \mathbf{E}_l - \mathbf{F}^{\mathrm T} \mathbf{E}^* \right)  \right. \nonumber \\
\left. + \varepsilon_0 J \left( \mathbf{F}^{-1} \mathbf{L}^{-1} \mathbf{F}^{- \mathrm T} + \mathbf{F}^{-1} \mathbf{L}^{-\mathrm T} \mathbf{F}^{- \mathrm T} \right) \left( \mathbf{E}_l - \mathbf{F}^{\mathrm T} \mathbf{E}^* \right) + \bdot{\mathbf{P}}_l \right\} = \bdot{\sigma}_E,
\end{align} 
\begin{align}
\mathbf{N} \times \left\{ J^{-1} \mu_0^{-1} \mathbf{c} \bdot{\mathbf{B}}_l - J^{-1} (\mbox{div} \, \mathbf{u}) \mu_0^{-1} \mathbf{cB}_l + J^{-1} \mu_0^{-1} \left( \mathbf{F}^{\mathrm T} \mathbf{LF} + \mathbf{F}^{\mathrm T} \mathbf{L}^{\mathrm T} \mathbf{F} \right) \mathbf{B}_l \right. \nonumber \\
- \bdot{\mathbf{M}}_l - \bdot{\mathbf{V}} \times \left( \varepsilon_0 J \mathbf{c}^{-1} \mathbf{E}_l + \mathbf{P}_l \right) - \varepsilon_0 \mathbf{V} \times \left( J (\mbox{div}\, \mathbf{u}) \mathbf{c}^{-1} \mathbf{E}_l + J \mathbf{c}^{-1} \bdot{\mathbf{E}}_l  \right. \nonumber \\
\left. \left. + J \left( \mathbf{F}^{-1} \mathbf{L}^{-1} \mathbf{F}^{- \mathrm T} + \mathbf{F}^{-1} \mathbf{L}^{-\mathrm T} \mathbf{F}^{- \mathrm T} \right) \mathbf{E}_l + \bdot{\mathbf{P}}_l \right) - \mu_0^{-1}  \mathbf{F}^{\mathrm T} \mathbf{L}^{\mathrm T} \mathbf{B}^* - \mu_0^{-1} \mathbf{F}^{\mathrm T} \bdot{\mathbf{B}}^* \right\} \nonumber \\
 = \bdot{\mathbf{K}}_l - \bdot{\sigma}_E \mathbf{V}_{\mathrm s} - \sigma_E \bdot{\mathbf{V}}_{\mathrm s}.
\end{align}

We recollect the Nanson's formula $\mathbf{n}\, da = J \mathbf{F}^{-\mathrm T} \mathbf{N}\, dA$ connecting reference and current area elements $dA$ and $da$, where $\mathbf{n}$ is the unit outward normal to $\partial \mathcal B_t$. The surface current and electric charge densities in the two configurations are related by $\mathbf{K}_l = \mathbf{F}^{-1} \mathbf{K} da/dA$ and $\sigma_E = \sigma_e da/dA$. Using these relations and the relations \eqref{pushed forward}, we rewrite the above incremental boundary conditions in the current configuration on $\partial \mathcal B_t$ as
\begin{equation} \label{B bc 1}
\mathbf{n} \times \left( \bdot{\mathbf{E}}_{l0} + (\bdot{\mathbf{v}} - \mathbf{Lv}) \times \mathbf{B} + \mathbf{v} \times \bdot{\mathbf{B}}_{l0} - \mathbf{LE}^* - \bdot{\mathbf{E}}^* \right) = \mathbf{0},
\end{equation}
\begin{equation} \label{B bc 2}
\mathbf{n} \cdot \left( \bdot{\mathbf{B}}_{l0} - (\mbox{div}\, \mathbf{u}) \mathbf{B}^* - \mathbf{L}^{-1} \mathbf{B}^* - \bdot{\mathbf{B}}^* \right) = 0,
\end{equation}
\begin{align}
\mathbf{n} \cdot \left\{ \varepsilon_0 \left( \bdot{\mathbf{E}}_{l0} - \mathbf{L}^{\mathrm T} \mathbf{E}^* - \bdot{\mathbf{E}}^* \right) + \varepsilon_0 (\mbox{div}\, \mathbf{u}) \left( \mathbf{E} -\mathbf{E}^* \right) + \bdot{\mathbf{P}}_{l0}  \right. \nonumber \\
+ \varepsilon_0 \left( \mathbf{L}^{-1} + \mathbf{L}^{-\mathrm T} \right) \left( \mathbf{E} - \mathbf{E}^* \right) = \bdot{\sigma}_{E0}, \label{B bc 3}
\end{align}
\begin{align}
\mathbf{n} \times \left\{ \mu_0^{-1} \bdot{\mathbf{B}}_{l0} - \mu_0^{-1} (\mbox{div}\, \mathbf{u}) \mathbf{B} + \mu_0^{-1} \left( \mathbf{L} + \mathbf{L}^{\mathrm T} \right) \mathbf{B} - \bdot{\mathbf{M}}_{l0} - \left( \bdot{\mathbf{v}} - \mathbf{Lv} \right) \times \left( \varepsilon_0 \mathbf{E} + \mathbf{P} \right) \right. \nonumber \\
\left. - \varepsilon_0 \mathbf{v} \times \left( (\mbox{div}\,\mathbf{u}) \mathbf{E} + \bdot{\mathbf{E}}_{l0} + \left( \mathbf{L}^{-1} + \mathbf{L}^{- \mathrm T} \right) \mathbf{E} + \bdot{\mathbf{P}}_{l0} \right) - \mu_0^{-1} \mathbf{L}^{\mathrm T} \mathbf{B}^* - \mu_0^{-1} \bdot{\mathbf{B}}^* \right\} \nonumber \\
= \bdot{\mathbf{K}}_{l0} + \mathbf{L}^{-1} \mathbf{K}  + \sigma_e \mathbf{Lv}_{\mathrm s}, \label{B bc 4}
\end{align}
where we have defined $\bdot{\mathbf{K}}_{l0} = \mathbf{F} \bdot{\mathbf{K}}_l \, dA/da$ and $\bdot{\sigma}_{E0} = \bdot{\sigma}_E \, dA/da$.

At the boundary $\partial \mathcal P$, incremental form of the boundary conditions \eqref{wire boundary} is given as
\begin{align}
\mathbf{n} \times \left( \bdot{\mathbf{E}} - \bdot{\mathbf{E}}^* \right) = \mathbf{0}, & \quad \mathbf{n} \cdot \left( \varepsilon_r \bdot{\mathbf{E}} - \bdot{\mathbf{E}}^* \right) = 0, \nonumber \\
\mathbf{n} \times \left( \frac{\bdot{\mathbf{B}}}{\mu_r} - \bdot{\mathbf{B}}^* \right) = \mathbf{0}, & \quad \mathbf{n} \cdot \left( \bdot{\mathbf{B}} - \bdot{\mathbf{B}}^* \right) = 0. \label{P bc}
\end{align}

Thus, a generic EMAT problem requires the solution of equations \eqref{ch2 up bal 1-2} and \eqref{ch6 general bal 3}--\eqref{ch6 general bal 7} in $\mathcal B_t$, equations \eqref{wire increm maxwell} in $\mathcal P$, and equations \eqref{vacuum increm maxwell} in vacuum using the boundary conditions \eqref{B bc 1}--\eqref{B bc 4} at $\partial \mathcal B_t$ and \eqref{P bc} at $\partial \mathcal P$.

\section{Concluding remarks}

This paper reviews the existing theory of electromagnetic interactions with a solid continuum. In particular, we analyze the problem of wave propagation in a finitely deformed elastic solid with an underlying electric, magnetic, and temperature field. Unlike the existing literature, we generalize a nonlinear theory of elasticity to include electromagnetic effects. On a finite deformation, magnetic, and electric field, the equations are linearized to consider wave propagation and several moduli tensors are introduced. 
In addition to filling a gap in literature, it is expected that this work will lead to an accurate mathematical modelling of the `Electromagnetic acoustic transducers' in particular.

\bigskip
\noindent \textbf{Acknowledgements:}

\noindent This work was undertaken when the author was at the University of Glasgow supported by a university postgraduate scholarship and a UK ORS scholarship. Thanks are also extended to Prof. Ray W. Ogden, FRS of the University of Glasgow, UK for his helpful guidance during the initial phase of this work.


\begin{thebibliography}{10}

\bibitem{Hirao2003}
M.~Hirao, H.~Ogi, EMATs for Science and Industry: Noncontacting Ultrasonic
  Measurements, Kluwer Academic Publishers, 2003.

\bibitem{Ludwig1993}
R.~Ludwig, Z.~You, R.~Palanisamy,
  Numerical
  simulations of an electromagnetic acoustic transducer-receiver system for NDT
  applications, IEEE Transactions on Magnetics 29(3) (1993) 2081--2089.


\bibitem{Ogi1997}
H.~Ogi, Field
  dependence of coupling efficiency between electromagnetic field and
  ultrasonic bulk waves, Journal of Applied Physics 82(8) (1997) 3940.

\bibitem{Thompson1978}
R.~B. Thompson, {A Model for the Electromagnetic Generation of Ultrasonic
  Guided Waves in Ferromagnetic Metal Polycrystals}, IEEE Transactions on
  Sonics and Ultrasonics 25(1) (1978) 7--15.

\bibitem{Shapoorabadi2005}
R.~J. Shapoorabadi, A.~Konrad, A.~N. Sinclair, The governing electrodynamic
  equations of electromagnetic acoustic transducers, Journal of Applied
  Physics 97(10) (2005) 6--8.


\bibitem{Pao1978}
Y.~H. Pao, Electromagnetic forces in deformable continua, in: S.~Nemat-Nasser
  (Ed.), Mechanics Today, Vol. 4, Oxford University Press, 1978, pp. 209--305.

\bibitem{Eringen1990}
A.~C. Eringen, G.~A. Maugin, Electrodynamics of Continua, Vol. 1,
  Springer-Verlag, 1990.

\bibitem{Maugin2009}
G.~A. Maugin, {On modelling electromagnetomechanical interactions in deformable
  solids}, International Journal of Advances in Engineering Sciences and
  Applied Mathematics 1 (2009) 25--32.

\bibitem{Jolly1996}
M.~R. Jolly, J.~D. Carlson, B.~C. Mu\~{n}oz, A model of the behaviour of
  magnetorheological materials, Smart Materials and Structures 5(5) (1996)
  607--614.

\bibitem{Bose2012}
H.~B\"{o}se, R.~Rabindranath, J.~Ehrlich, Soft magnetorheological elastomers
  as new actuators for valves, Journal of Intelligent Material Systems and
  Structures 23(9) (2012) 989--994.

\bibitem{Dorfmann2004}
A.~Dorfmann, R.~W. Ogden, Nonlinear magnetoelastic deformations, Quarterly
  Journal of Mechanics and Applied Mathematics 57(7) (2004) 599--622.

\bibitem{Dorfmann2006}
A.~Dorfmann, R.~W. Ogden, {Nonlinear electroelastic deformations}, Journal of
  Elasticity 82(2) (2006) 99--127.

\bibitem{Destrade2011}
M.~Destrade, R.~W. Ogden, On magneto-acoustic waves in finitely deformed elastic solids, Mathematics
and Mechanics of Solids 16(6) (2011) 594--604.

\bibitem{Saxena2011}
P.~Saxena, R.~W. Ogden, On surface
  waves in a finitely deformed magnetoelastic half-space, International
  Journal of Applied Mechanics 3(4) (2011) 633--665.

\bibitem{Saxena2012}
P.~Saxena, On Wave Propagation in
  Finitely Deformed Magnetoelastic Solids, Ph.D. thesis, University of
  Glasgow (2012).
	
	\bibitem{Coleman1963}
	B.~D.~Coleman, W.~Noll, The thermodynamics of elastic materials with heat conduction and viscosity, Archive for Rational Mechanics and Analysis 13(1) (1963) 167--178.

\end{thebibliography}
\end{document}